\def\uma{\rm 1\!\!\hskip 1 pt l}

\documentclass[twocolumn,showpacs,showkeys,aps]{revtex4} 
\usepackage{amsmath}
\usepackage{amsfonts}
\usepackage{amssymb}

\begin{document}

\title{Non-Relativistic Propagators via Schwinger's Method}

\author{A. Arag\~{a}o} \email{aharagao@if.ufrj.br}
\affiliation{Instituto de F\'{\i}sica, Universidade Federal do Rio de Janeiro,
Caixa Postal 68.528, 21941-972 Rio de Janeiro, RJ, Brazil.}
%
\author{F. A. Barone}
\email{fbarone@unifei.edu.br}
\affiliation{Universidade Federal de Itajub\'a,
Av. BPS 1303\\  Caixa Postal 50 - 37500-903, Itajub\'a, MG, Brazil.}
\author{H. Boschi-Filho}\email{boschi@if.ufrj.br}
\affiliation{Instituto de F\'{\i}sica, Universidade Federal do Rio de Janeiro,
Caixa Postal 68.528, 21941-972 Rio de Janeiro, RJ, Brazil.}
\author{C. Farina}\email{farina@if.ufrj.br}
\affiliation{Instituto de F\'{\i}sica, Universidade Federal do Rio de Janeiro,
Caixa Postal 68.528, 21941-972 Rio de Janeiro, RJ, Brazil.}

\begin{abstract}
In order to popularize the so called Schwinger's method we
reconsider the Feynman propagator of two non-relativistic systems: a
charged particle in a uniform magnetic field and a charged harmonic
oscillator in a uniform magnetic field. Instead of solving the
Heisenberg equations for the position  and the canonical momentum
operators, ${\bf R}$ and ${\bf P}$, we apply this method by solving
the Heisenberg equations for the gauge invariant operators ${\bf R}$
and $\mbox{\mathversion{bold}${\pi}$} = {\bf P}-e{\bf A}$, the
latter being the mechanical momentum operator. In our procedure we
avoid fixing the gauge from the beginning and the result thus obtained
shows explicitly the gauge dependence of the Feynman propagator.
\end{abstract}

 \maketitle

\section{Introduction}

In a recent paper published in this journal
\cite{BaroneBoschiFarinaAJP2003}, three methods were used to compute
the Feynman propagators of a one-dimensional harmonic oscillator, with the purpose of
allowing a student to compare the advantadges and disadvantadges of
each method. The above mentioned methods were the following: the so called
Schwinger's method (SM), the algebraic method and the path integral
one. Though extremely powerful and elegant, Schwinger's method is by
far the less popular among them. The main purpose of the present
paper is to popularize Schwinger's method providing the reader with
two examples slightly more difficult than the harmonic oscillator
case and whose solutions may serve as a preparation for attacking
relativistic problems. In some sense, this paper is complementary to
reference \cite{BaroneBoschiFarinaAJP2003}.

The method we shall be concerned with was introduced by Schwinger in
1951 \cite{Schwinger1951} in a paper about QED entitled \lq\lq Gauge
invariance and vacuum polarization". After introducing the proper
time representation for computing effetive actions in QED, Schwinger
was faced with a kind of non-relativistic propagator in one extra
dimension. The way he solved this problem is what we mean by
Schwinger's method for computing quantum propagators.
 For relativistic Green functions of charged particles under external
electromagnetic fields, the main steps of this method are summarized
in Itzykson and Zuber's textbook \cite{ItziksonZuberBook} (apart, of
course, from Schwinger's work \cite{Schwinger1951}). Since then, this
method has been used mainly in relativistic quantum theory
\cite{GitmanBook,Dodonov75,Dodonov76,Lykken,Ferrando:1994vt,BFV96,Gavrilov:1998hw,
McKeon:1998zx,Chyi:1999fc,Tsamis:2000ah,Chaichian:2000eh,Chung:2001mb,BoschiFarinaVaidya1996}.

However, as mentioned before, Schwinger's method is also well suited
for computing non-relativistic propagators, though it has rarely
been used in this context. As far as we know, this method was used
for the first time in non-relativistic quantum mechanics by Urrutia and
Hernandez \cite{UrrutiaHernandez1984}. These authors used
Schwinger's action principle to obtain the Feynman propagator for a
damped harmonic oscillator with a time-dependent frequency under a
time-dependent external force. Up to our knowledge, since then only
a few papers have been written with this method, namely: in 1986,
Urrutia and Manterola \cite{UrrutiaManterola1986} used it in the
problem of an anharmonic charged oscillator under a magnetic field;
in the same year, Horing, Cui, and Fiorenza
\cite{HoringCuiFiorenza1986} applied Schwinger's method to obtain
the Green function for crossed time-dependent electric and
magnetic fields; the method was later applied in a rederivation of
the Feynman propagator for a harmonic oscillator with a
time-dependent frequency \cite{FarinaSegui1993}; a connection with
the mid-point-rule for path integrals involving electromagnetic
interactions was discussed in \cite{RabelloFarina1995}. Finally, pedagogical
presentations of this method can be found in the recent publication
\cite{BaroneBoschiFarinaAJP2003} as well as in
Schwinger's original lecture notes recently published
\cite{SchwingerBookEnglert2001}, which includes a discussion of the
quantum action principle and a derivation of the method to calculate
propagators with some examples.


It is worth mentioning  that this same method was independently
developed by M. Goldberger and M. GellMann in the autumn of 1951  in
connection with an unpublished paper about density matrix in
statistical mechanics \cite{Goldberger1951}.

Our purpose in this paper is to provide the reader with two other
examples of non-relativistic quantum propagators that can be
computed in a straightforward way by Schwinger's method, namely: the
propagator for a charged particle in a uniform magnetic field and
this same problem with an additional harmonic oscillator potential.
Though these problems have already been treated in the context of the quantum
action principle \cite{UrrutiaManterola1986}, we decided to reconsider them for the
following reasons: instead of solving the
Heisenberg equations for the position  and the canonical momentum
operators, ${\bf R}$ and ${\bf P}$, as is done in \cite{UrrutiaManterola1986},
we apply Schwinger's method by solving the Heisenberg equations for the gauge invariant operators ${\bf R}$
and $\mbox{\mathversion{bold}${\pi}$} = {\bf P}-e{\bf A}$, the
latter being the mechanical momentum operator. This is precisely the procedure followed by
Schwinger in his seminal paper of gauge invariance and vacuum polarization \cite{Schwinger1951}.
This procedures has some nice properties. For instance,  we are not obligued to choose a particular
gauge at the beginning of calculations. As a consequence, we end up with an expression
for the propagator written in an arbitrary gauge.  As a bonus, the transformation law
for the propagator under gauge transformations can be readly obtained.

In order to prepare the students to attack  more complex problems,
we solve the Heisenberg equations in matrix form, which is well suited for
generalizations involving Green functions of relativistic charged particles
under the influence of electromagnetic fields (constant $F_{\mu\nu}$,
a plane wave field or even combinations of both). For pedagogical reasons, at the end of each
calculation, we show how to extract the corresponding energy spectrum from the Feynman propagator.
Although the way Schwniger's method must be applied to non-relativistic problems has already been
 explained in the literature
 \cite{UrrutiaManterola1986,SchwingerBookEnglert2001,BaroneBoschiFarinaAJP2003},
 it is not of common knowledge so that we start this paper by summarizing its main steps.
 The paper is organized as follows: in the next section we review Schwinger's method, in section
 \ref{SectionExamples} we present our examples and section \ref{SectionFinalRemarks} is left for the
 final remarks.

\section{Main steps of Schwinger's method}

For simplicity, consider a one-dimensional time-independent Hamiltonian $\mathcal{H}$ and the
corresponding non-relativistic Feynman propagator defined as
\begin{equation}\label{DefinicaoPropagador}
K(x,x^{\prime};\tau)=\theta(\tau) \langle x | \exp{ \Big[ {\frac{-i
\cal{H} \tau}{\hbar}} \Big]} | x^{\prime} \rangle,
\end{equation}
where $\theta(\tau)$ is the Heaviside step function and $|x\rangle$,
$|x^{\prime}\rangle$ are the eingenkets of the position operator $X$
(in the Schr\"odinger picture) with eingenvalues $x$ and
$x^{\prime}$, respectively. The extension for 3D systems is
straightforward and will be done in the next section. For $\tau>0$
we have, from equation (\ref{DefinicaoPropagador}), that
\begin{equation} \label{Eq2}
i\hbar\frac{\partial}{\partial\tau}K(x,x^{\prime};\tau) =
 \langle x |\mathcal{H} \exp{\Big[ {\frac{-i \mathcal{H} \tau}{\hbar}}\Big]} |
x^{\prime}\rangle.
\end{equation}
Inserting  the unity $\uma =\exp{[-(i/\hbar)\mathcal{H}\tau]}\exp{[(i/\hbar)\mathcal{H}\tau]}$
 in the r.h.s. of the above expression and using the well known relation
  between operators in the Heisenberg and Schr\"odinger pictures,
   we get the equation for the Feynman propagator in the Heisenberg picture,
\begin{equation}
i\hbar\frac{\partial}{\partial\tau}
 K(x,x^{\prime};\tau)=\langle x,\tau |\mathcal{H} (X(0),P(0))| x^{\prime},0\rangle, \label{Eq3}
\end{equation}
where $|x,\tau\rangle$ and $|x^{\prime},0\rangle$ are the
eingenvectors of  operators $X(\tau)$ and $X(0)$, respectively, with
the corresponding eingenvalues $x$ and $x^{\prime}$:
$X(\tau)|x,\tau\rangle=x|x,\tau\rangle$ and
$X(0)|x^{\prime},0\rangle=x^{\prime}|x^{\prime},0\rangle$, with
$K(x,x^{\prime};\tau)=\langle x,\tau|x^{\prime},0\rangle$. Besides,
$X(\tau)$ and $P(\tau)$ satisfy the Heisenberg equations,
\begin{equation}
i\hbar \frac{dX}{d\tau}(\tau)=[X(\tau),\mathcal{H}] \hspace{0.2cm} ;
 \hspace{0.2cm}  i\hbar \frac{dP}{d\tau}(\tau)=[P(\tau),\mathcal{H}].
\label{Eq4}
\end{equation}
Schwinger's method consists in the following steps:
\begin{description}
\item[(i)] we solve the Heisenberg equations for $X(\tau)$ and $P(\tau)$, and write
the solution for $P(0)$ only in terms of the operators $X(\tau)$ and $X(0)$;
\item[(ii)] then, we substitute the results obtained in {\bf (i)} into the expression
for $\mathcal{H} (X(0),P(0))$ in  (\ref{Eq3}) and using the
commutator $[X(0),X(\tau)]$ we rewrite each term of $\mathcal{H}$ in
a time ordered form with all operators $X(\tau)$ to the left and all
operators $X(0)$ to the right;
\item[(iii)] with such an ordered hamiltonian, equation (\ref{Eq3}) can be readly cast into the form
\begin{equation}\label{Eq5}
i\hbar\frac{\partial}{\partial\tau} K(x,x^{\prime};\tau) =
 F(x,x^{\prime};\tau)K(x,x^{\prime};\tau),
\end{equation}
with $F(x,x^{\prime};\tau)$ being an ordinary function defined as
\begin{equation}\label{Eq6}
F(x,x^{\prime};\tau)=\frac{\langle x,\tau |\mathcal{H}_{ord}
(X(\tau),X(0))| x^{\prime},0\rangle}
 {\langle x,\tau|x^{\prime},0\rangle}.
\end{equation}
Integrating in $\tau$, the Feynman propagator takes the form
\begin{equation}\label{Eq7}
K(x,x^{\prime};\tau)=C(x,x^{\prime})
 \exp\left\{\!\!-\frac{i}{\hbar}\!\! \int^{\tau}
  \!\!\!\!\! F(x,x^{\prime};\tau^\prime)d\tau^{\prime}\!\right\} ,
\end{equation}
where $C(x,x^{\prime})$ is an integration constant independent of $\tau$ and $\int^{\tau}$
means an indefinite integral;
\item[(iv)] last step is concerned with the evaluation of $C(x,x^{\prime})$.
 This is done after imposing the following conditions
\begin{eqnarray}\label{CondicaoP(tau)}
-i\hbar\frac{\partial}{\partial x}\langle
 x,\tau|x^{\prime},0\rangle &=&
 \langle x,\tau|P(\tau)|x^{\prime},0\rangle \, ,  \\
i\hbar\frac{\partial}{\partial x^{\,\prime}}\langle
x,\tau|x^{\prime},0\rangle &=&
 \langle x,\tau|P(0)|x^{\prime},0\rangle\, , \label{CondicaoP(0)}
\end{eqnarray}
 as well as the initial condition
\begin{equation}\label{CondicaoInicial}
\lim_{\tau\rightarrow 0^+}
 K(x,x^{\prime};\tau) = \delta(x-x^{\prime})\ .
\end{equation}
\end{description}

Imposing conditions (\ref{CondicaoP(tau)}) and (\ref{CondicaoP(0)})
means to substitute in their left hand sides  the expression for
$\langle x,\tau|x^{\prime},0\rangle$ given by (\ref{Eq7}), while in
their right hand sides the operators $P(\tau)$ and $P(0)$,
respectively, written in terms of the operators $X(\tau)$ and $X(0)$
with the appropriate time ordering.


\section{Examples}\label{SectionExamples}

\subsection{Charged particle in an uniform magnetic field}
\label{cpumf}

As our first example, we consider the propagator of a
non-relativistic particle with electric charge $e$ and mass $m$,
submitted to a constant and uniform magnetic field ${\bf B}$. Even
though this is a genuine three-dimensional problem, the extension of
the results reviewed in the last section to this case is
straightforward. Since there is no electric field present, the
hamiltonian can be written as
\begin{equation}\label{HamiltonianaBUniforme}
\mathcal{H}=\frac{\left({\bf P} - e{\bf A}\right)^{2}}{2m}
 = \frac{\mbox{\mathversion{bold}${\pi}$}^2}{2m}\ ,
\end{equation}
where ${\bf P}$  is the canonical momentum operator,
 ${\bf A}$  is the vector potential and
 $\mbox{\mathversion{bold}${\pi}$}={\bf P}-e{\bf A}$ is the gauge
 invariant mechanical momentum operation. We choose the axis such that the magnetic
 field is given by ${\bf B}=B{\bf e_3}$. Hence, the hamiltonian can be decomposed as
\begin{equation}
\mathcal{H} = \frac{\pi_{1}^{2}+\pi_{2}^{2}}{2m} +
 \frac{P_3^2}{2m} \; =\; \mathcal{H}_{\bot} +
 \frac{P_{3}^{2}}{2m}\ ,
\end{equation}
with an obvious definition for $\mathcal{H}_{\bot}$.

Since the motion along the ${\cal OX}_3$ direction is free, the
three-dimensional propagator $K({\bf x},{\bf x}^{\prime};\tau)$ can
be written as a product of a two-dimensional propagator,
$K_{\bot}({\bf r},{\bf r}^{\prime};\tau)$, related to the magnetic
field and a one-dimensional free propagator,
$K_{3}^{(0)}(x_3,x_3^{\prime};\tau)$:
\begin{equation}\label{DecomposicaoPropagador}
K({\bf x},{\bf x}^{\prime};\tau)=K_{\bot}({\bf r},{\bf
r}^{\prime};\tau)
 K_{3}^{(0)}(x_3,x_3^{\prime};\tau),\;\;\;\;
(\tau>0)
\end{equation}
where ${\bf r} = x_1{\bf e_1} + x_2{\bf e_2} $ and
$K_{3}^{(0)}(x_3,x_3^{\prime};\tau)$ is the well known propagator of
the free particle \cite{FeynmanHibbsBook},
\begin{equation}
K_{3}^{(0)}(x_3,x_3^{\prime};\tau)=\sqrt{\frac{m}{2\pi i\hbar \tau}}
 \exp{ \Big[ \frac{im}{2\hbar}\frac{(x_3 - x_3^{\prime})^{2}}{\tau}\Big]}.
\label{Eq14}
\end{equation}
 In order to use Schwinger's method to compute the two-dimensional
 propagator $K_{\bot}({\bf r},{\bf r}^{\prime};\tau) =
  \langle{\bf r},\tau|{\bf r}^{\prime},0\rangle$, we start by
 writing the differential equation
\begin{equation}\label{EqDiferencialPropagador2D}
i\hbar\frac{\partial}{\partial\tau}
 \langle{\bf r},\tau|{\bf r}^{\prime},0\rangle =
 \langle {\bf r},\tau|\mathcal{H}_{\bot}({\bf R}_\bot(0),
 \mbox{\mathversion{bold}${\pi}$}_\bot(0)  )| {\bf r}^{\prime},0\rangle\ ,
\end{equation}
where ${\bf R}_\bot(\tau) =
 X_{1}(\tau){\bf e_1} + X_{2}(\tau){\bf e_2}$ and
 $\mbox{\mathversion{bold}${\pi}$}_\bot(\tau) =
  \pi_{1}(\tau){\bf e_1} + \pi_{2}(\tau){\bf e_2}$.
In (\ref{EqDiferencialPropagador2D})
 $|{\bf r},\tau\rangle$ and $|{\bf r}^{\prime},0\rangle$ are the
eigenvectors of position operators ${\bf R}(\tau)=X_{1}(\tau){\bf
e_1} + X_{2}(\tau){\bf e_2}$ and ${\bf R}(0)=X_{1}(0){\bf e_1} +
X_{2}(0){\bf e_2}$, respectively. More especifically, operators
$X_{1}(0)$, $X_{1}(\tau)$, $X_{2}(0)$ and $X_{2}(\tau)$ have the
eigenvalues $x_{1}^{\prime}$, $x_{1}$, $x_{2}^{\prime}$ and $x_{2}$,
respectively.

\noindent
 In order to solve the Heisenberg equations for operators
 ${\bf R}_\bot(\tau)$ and
 $\mbox{\mathversion{bold}${\pi}$}_\bot(\tau)$, we need the
 commutators
\begin{eqnarray}\label{ComutadorXcomPi2}
\Big[X_{i}(\tau) , \pi_{j}^{2}(\tau) \Big] &=&
 2i\hbar \pi_{i}(\tau)\, ,\cr
  \Big[\pi_{i}(\tau) , \pi_{j}^{2}(\tau)\Big] &=& 2i\hbar eB\epsilon_{ij3}\pi_{j}(\tau),
  \label{ComutadorPicomPi}
\end{eqnarray}
where $\epsilon_{ij3}$ is the usual Levi-Civita symbol. Introducing
the matrix notation
\begin{equation}
{\bf R}(\tau)=\left(
\begin{array}{c}
X_{1}(\tau)   \\ X_{2}(\tau)
\end{array}
\right)
\hspace{0.25cm};\hspace{0.25cm}
\mbox{\mathversion{bold}${\Pi}$}(\tau)=\left(
\begin{array}{c}
\pi_{1}(\tau)   \\ \pi_{2}(\tau)
\end{array}
\right)
\label{Eq20}\ ,
\end{equation}
and using the previous commutators the Heisenberg equations of
motion can be cast into the form
\begin{eqnarray}\label{EquacaoR}
\frac{d{\bf R}(\tau)}{d\tau}&=&\frac{\mbox{\mathversion{bold}${\Pi}$}(\tau)}{m}\ , \\
 \frac{d\mbox{\mathversion{bold}${\Pi}$}(\tau)}{d\tau} &=& 2\omega
\mathbb{C}  \mbox{\mathversion{bold}${\Pi}$}(\tau) \,,
 \label{EquacaoPi}
\end{eqnarray}
where $2\omega={eB}/{m}$ is the cyclotron frequency and we defined the anti-diagonal matrix
\begin{equation}
\mathbb{C}=\left(
\begin{array}{cc}
 0 & 1 \\
-1 & 0
\end{array}
\right)
\label{Eq21}\ .
\end{equation}
Integrating equation (\ref{EquacaoPi}) we find
\begin{eqnarray}
\mbox{\mathversion{bold}${\Pi}$}(\tau) &=&
 e^{2\omega\mathbb{C}\tau}\mbox{\mathversion{bold}${\Pi}$}(0)
\label{Eq23}\ .
\end{eqnarray}
Substituting this solution in equation (\ref{EquacaoR}) and
integrating once more, we get
\begin{eqnarray}\label{Eq22}
\textbf{R}(\tau)-\textbf{R}(0)&=&\frac{\sin{(\omega\tau)}}{m\omega}
 e^{\omega\mathbb{C}\tau}\mbox{\mathversion{bold}${\Pi}$}(0)
\  ,
\end{eqnarray}
where we used the following properties of $\mathbb{C}$ matrix:
 \linebreak
 $\mathbb{C}^{2}\!=\!-\uma$;
$\mathbb{C}^{-1}\!=\!-\mathbb{C}=\mathbb{C}^{T}$,
 $e^{\alpha\mathbb{C}} =
 \cos{(\alpha)}\uma + \sin{(\alpha)}\mathbb{C}$
with $\mathbb{C}^{T}$ being the transpose of $\mathbb{C}$. Combining
equations (\ref{Eq22}) and (\ref{Eq23}) we can write
$\mbox{\mathversion{bold}${\Pi}$}(0)$ in terms of the operators
${\bf R}(\tau)$ and ${\bf R}(0)$ as
\begin{equation}\label{Eq24}
\mbox{\mathversion{bold}${\Pi}$}(0)=\frac{m\omega}{\sin{(\omega\tau)}}
 e^{-\omega\mathbb{C}\tau}\biggl({\bf R}(\tau)-{\bf R}(0)\biggr).
\end{equation}
In order to express $\mathcal{H}_\bot = (\pi_1^2 + \pi_2^2)/2m$ in
terms of $\textbf{R}(\tau)$ and $\textbf{R}(0)$, we use
(\ref{Eq24}). In matrix notation, we have
\begin{eqnarray}
\mathcal{H}_{\bot}&=&\frac{1}{2m}\,\mbox{\mathversion{bold}${\Pi}$}^{T}(0)\mbox{\mathversion{bold}${\Pi}$}(0)\nonumber\\
&=&\frac{m\omega^2}{2\sin^2{(\omega\tau)}}\biggl({\bf R}^{T}(\tau){\bf R}(\tau)+{\bf R}^{T}(0){\bf R}(0)+\nonumber\\
&\ &\hspace{1.5cm}-{\bf R}^{T}(\tau){\bf R}(0)-{\bf R}^{T}(0){\bf R}(\tau)\biggr)\ .
\label{Hnaoordenado}
\end{eqnarray}
Last term on the r.h.s. of (\ref{Hnaoordenado}) is not ordered
appropriately as required in the step ({\bf ii}). The correct
ordering may be obtained as follows: first, we write
\begin{equation}
\textbf{R}(0)^{T}  \textbf{R}(\tau)
 = \textbf{R}(\tau)^{T}\textbf{R}(0) +
\sum_{i=1}^{2}[X_{i}(0) , X_{i}(\tau)]\, .
\end{equation}
Using equation (\ref{Eq22}), the usual commutator
$[X_{i}(0),{\pi}_j(0)]=i\hbar\delta_{ij}\uma$ and the properties of
matrix $\mathbb{C}$ it is easy to show that
\begin{equation}\label{Eq25}
\sum_{i=1}^{2}[X_{i}(0) , X_{i}(\tau)]
 =
\frac{2i\hbar\sin(\omega\tau)\cos(\omega\tau)}{m\omega}\, ,
\end{equation}
so that hamiltonian $\mathcal{H}_{\bot}$ with the appropriate time
ordering takes the form
\begin{eqnarray}
\mathcal{H}_\perp&=&\frac{m\omega^2}{2\sin^2{(\omega\tau)}}
\biggl\{{\bf R}^{2}(\tau)+{\bf R}^{2}(0)-2{\bf R}^{T}(\tau){\bf R}(0)\biggr\} \, \nonumber \\ &-& i
 \hbar \omega\cot(\omega\tau).
\label{Eq27}
\end{eqnarray}
Substituting this hamiltonian into equation
(\ref{EqDiferencialPropagador2D}) and integrating in $\tau$, we
obtain
\begin{equation}\label{ProtoPropagador}
\langle{\bf r},\tau\vert{\bf r}^{\prime},0\rangle =
 \frac{C({\bf r},{\bf r}^{\prime})}{\sin{(\omega\tau)}}
\exp\biggl\{{\frac{im\omega}{ 2\hbar}}\cot(\omega\tau)({\bf r}-{\bf
r}^{\prime})^2\biggr\} ,
\end{equation}
where $C({\bf r},{\bf r}^{\;\prime})$ is an integration constant to
be determined by conditions (\ref{CondicaoP(tau)}),
(\ref{CondicaoP(0)}) and (\ref{CondicaoInicial}), which for the case
of hand read
\begin{eqnarray}
\langle{\bf r},\tau\vert\pi_j(\tau)\vert{\bf r}^{\prime},0\rangle
\!\!&=&\!\!
 \left(-i\hbar\frac{\partial}{\partial x_j}-eA_j({\bf r})\right)
\langle{\bf r},\tau\vert{\bf r}^{\prime},0\rangle
\label{Eq30}\\
\langle{\bf r},\tau\vert\pi_j(0)\vert{\bf r}^{\prime},0\rangle
 \!\!&=&\!\!
\left(i\hbar\frac{\partial}{\partial x^\prime_j}-eA_j({\bf
r}^{\prime})\right) \langle{\bf r},\tau\vert{\bf
r}^{\prime},0\rangle\ ,
\label{Eq31}\\
\lim_{\tau\rightarrow 0^+}\langle{\bf r},
 \tau\vert{\bf r}^{\prime},0\rangle
 \!\!&=&\!\! \delta^{(2)}({\bf r}-{\bf
r}^{\prime}).\hspace{3.45cm} \label{Eq32}
\end{eqnarray}

In order to compute the matrix element on the l.h.s. of
(\ref{Eq30}), we need to express
$\mbox{{\mathversion{bold}${\Pi}$}}(\tau)$ in terms of
 ${\bf R}(\tau)$ and ${\bf R}(0)$. From equaitons (\ref{Eq23}) and
(\ref{Eq24}), we have
\begin{equation}
\mbox{{\mathversion{bold}${\Pi}$}}(\tau)=
\frac{m\omega}{\sin{(\omega\tau)}}\mbox{\large
$e^{\omega\tau\mathbb{C}}$} \biggl({\bf R}(\tau)-{\bf R}(0)\biggr),
\label{Eq33}
\end{equation}
which leads to the matrix element
\begin{eqnarray}
\langle{\bf r},\tau\vert\pi_j(\tau)\vert{\bf r}^{\prime},0\rangle
 &=&
 m\omega[\cot(\omega\tau) \left(x_j-x^\prime_j \right) \nonumber\\
&+& \mbox{\large $\epsilon_{jk3}$}\left(x_k-x^\prime_k\right)]
\langle{\bf r},\tau\vert{\bf r}^{\prime},0\rangle\ , \label{Eq34a}
\end{eqnarray}
where we used the properties of matrix $\mathbb{C}$ and Einstein
convention for repeated indices is summed. Analogously, the l.h.s.
of equation (\ref{Eq31}) can be computed from (\ref{Eq24}),
\begin{eqnarray}
\label{Eq34b}
\langle{\bf r},\tau\vert\pi_j(0)\vert{\bf r}^{\prime},0\rangle &=& m\omega[\cot(\omega\tau) \left(x_j-x^\prime_j \right) \nonumber\\
&-&\mbox{\large $\epsilon_{jk3}$}\left(x_k-x^\prime_k\right)] \langle{\bf r},\tau\vert{\bf r}^{\prime},0\rangle\ .
\end{eqnarray}

Substituting equations (\ref{Eq34a}) and (\ref{Eq34b}) into
(\ref{Eq30}) and (\ref{Eq31}), respectively, and using
(\ref{ProtoPropagador}), we have
\begin{eqnarray}\label{Equacao1ParaC}
\Big[i\hbar\frac{\partial }{\partial x_j} +
 eA_j({\bf r})\! + {1\over 2} e
F_{jk}(x_k \!-\! x^\prime_k)\Big] C({\bf r},{\bf r}^{\;\prime})\! \!\!&=&\!\! 0,\\
\Big[i\hbar\frac{\partial }{\partial x^\prime_j} -eA_j({\bf
r}^{\;\prime}) \! + {1\over 2} e F_{jk}(x_k \! -\!x^\prime_k)\Big]\,
 \!C({\bf r},{\bf r}^{\;\prime}) \!\!\!&=& \!\!0,\label{Equacao2ParaC}
\end{eqnarray}
where we defined $F_{jk}={\mbox{\large $\epsilon$}}_{jk3}\, B$.

Our strategy to solve the above system of differential equations is
the following: we first equation (\ref{Equacao1ParaC}) assuming in
this equation variables ${{\bf r}}^{\;\prime}$ as constants. Then,
we impose that the result thus obtained is a solution of equation
(\ref{Equacao2ParaC}). With this goal, we multiply both sides of
(\ref{Equacao1ParaC}) by $dx_j$ and sum over $j$, to obtain
\begin{equation}\label{dlnC}
{1\over C}\left({\partial C\over\partial x_j}\; dx_j \right) =
{ie\over\hbar} \biggl[ A_j({\bf r})+{1\over 2}F_{jk}\left(
x_k-x^\prime_k\right)\biggr] \; dx_j\; .
\end{equation}
Integration of the previous equation leads to
\begin{equation}\label{integracaoCrr'1}
C({\bf r},{\bf r}^{\;\prime}) = C({{\bf r}}^{\;\prime},{{\bf
r}}^{\;\prime})\;
 \mbox{\Large$ e^{\{ {ie\over \hbar}_{\;\;\Gamma}
\int_{\;\;\atop{{{\bf r}}^{\;\prime}}}^{\;{\bf r}}
 [ A_j (\mbox{\footnotesize{\mathversion{bold}${\xi}$}}) +
{1\over 2}\, F_{jk}\left(\xi_k-x^\prime_k\right)] \; d\xi_j\}}$}\; ,
\end{equation}
where the line integral is assumed to be along curve $\Gamma$, to be
specified in a moment. As we shall see, this line integral does not
depend on the curve $\Gamma$ joining ${{\bf r}}^{\;\prime}$ and
${\bf r}$, as expected, since the l.h.s. of (\ref{dlnC}) is an exact
differencial.

In order to determine the differential equation for $C({{\bf
r}}^{\;\prime},{{\bf r}}^{\;\prime})$ we must substitue expression
 (\ref{integracaoCrr'1}) into equation (\ref{Equacao2ParaC}). Doing
 that and using carefully the fundamental theorem of differential
 calculus, it is straightforward to show that
\begin{equation}\label{integracaoCrr'2}
{\partial C\over\partial x^\prime_j}({{\bf r}}^{\;\prime},{{\bf
r}}^{\;\prime})=0\; ,
\end{equation}
which means that $C({\bf r}^{\,\prime},{{\bf r}}^{\;\prime})$  is a
constant, $C_0$, independent of ${{\bf r}}^{\;\prime}$. Noting that
\begin{equation}
[{\bf B}\times\left(\mbox{{\mathversion{bold}${\xi}$}}- {{\bf
r}}^{\;\prime}\right)]_j =
 -F_{jk}\left(\xi_k-x^\prime_k\right)\, ,
\end{equation}
equation (\ref{integracaoCrr'1}) can be written as
\begin{equation}\label{Cintegrado1}
C({\bf r},{{\bf r}}^{\;\prime}) \!= C_0\; \exp\left\{ {ie\over
\hbar}_{\;\;\Gamma}\!\! \int_{\;\;\atop{{{\bf
r}}^{\;\prime}}}^{\;{\bf r}}\!\! \bigl[{\bf A}
(\mbox{{\mathversion{bold}${\xi}$}})-{1\over 2}\, {\bf B}\times\left
(\mbox{{\mathversion{bold}${\xi}$}}-{{\bf
r}}^{\;\prime}\right)\bigr]\! \cdot\!
d\mbox{{\mathversion{bold}${\xi}$}}\right\} .
\end{equation}
Observe, now, that the integrand in the previous equation has a
vanishing curl,
$$
\mbox{{\mathversion{bold}${\nabla}$}}_{\mbox{{\mathversion{bold}${\xi}$}}}
\times\biggl[{\bf A} (\mbox{{\mathversion{bold}${\xi}$}})-{1\over
2}\, {\bf B} \times\left(\mbox{{\mathversion{bold}${\xi}$}}-{{\bf
r}}^{\;\prime}\right)\biggr]= {\bf B}-{\bf B}={\bf 0}\; ,
$$
which means that the line integral in (\ref{Cintegrado1}) is path
independent. Choosing, for convenience, the straightline from
 ${{\bf r}}^{\;\prime}$ to ${\bf r}$, it can be readly shown that
$$
{\;}_{\;\;\;\atop{\mbox{$\Gamma_{sl}$}}}\!\int_{\;\;\atop{{{\bf
r}}^{\;\prime}}}^{\;{\bf r}} [{\bf B}\times\left(
\mbox{{\mathversion{bold}${\xi}$}}-{{\bf r}}^{\;\prime} \right)]
\cdot d\mbox{{\mathversion{bold}${\xi}$}}=0\; ,
$$
where $\Gamma_{sl}$ means a straightline from
 ${{\bf r}}^{\;\prime}$ to ${\bf r}$. With this simplification,
 the  $C({{\bf r}}^{\;\prime},{\bf r})$ takes the form
\begin{equation}\label{Cintegrado1}
C({\bf r},{{\bf r}}^{\;\prime}) \!= C_0\; \exp\left\{ {ie\over
\hbar}_{\;\;\Gamma_{sl}}\!\! \int_{\;\;\atop{{{\bf
r}}^{\;\prime}}}^{\;{\bf r}}\!\! {\bf A}
(\mbox{{\mathversion{bold}${\xi}$}}) \cdot
d\mbox{{\mathversion{bold}${\xi}$}}\right\} .
\end{equation}
Substituting last equation into (\ref{ProtoPropagador}) and using
the initial condition (\ref{CondicaoInicial}), we readly obtain
$C_{0}=\frac{m\omega}{2\pi i\hbar}$. Therefore the complete Feynman
propagator for a charged particle under the influence of a constant
and uniform magnetic field takes the form
\begin{eqnarray}\label{PropagadorFinalBUniforme}
K({\bf x},{{\bf x}}^{\prime};\tau)\hspace{6.0cm}\nonumber\\
={m\, \omega\over 2\pi i\hbar\,\sin{(\omega\tau)}}  \sqrt{{m\over 2\pi i\hbar\tau}}
\exp\left\{ {ie\over \hbar}
 \int_{{\bf r}^{\prime}}^{{\bf r}}\!\!\!\!\!{\bf A}
  (\mbox {{\mathversion {bold}${\xi}$}}) \cdot d\mbox{{\mathversion{bold}${\xi}$}}\right\}\nonumber\\
\exp\biggl\{{im\omega\over 2\hbar}\cot(\omega\tau)({\bf r}-{{\bf r}}^{\;\prime})^2\biggr\}
 \exp\biggl\{{im\over 2\hbar}{\left( x_3 - x_3^\prime\right)^2\over
 \tau}\biggr\}\, ,
\end{eqnarray}
where in the above equation we omitted the symbol $\Gamma_{sl}$ but,
of course, it is implicit that the line integral must be done along
a straightline, and we brought back the free propagation along the
${\cal OX}_3$ direction. A few comments about the above result are
in order.
\begin{enumerate}
\item Firstly, we should emphasize that the line integral which
appears in the first exponencial on the r.h.s. of
 (\ref{PropagadorFinalBUniforme}) must be evaluated along a straight
 line between ${\bf r}^{\prime}$ and ${\bf r}$. If for some reason
 we want to choose another path, instead of integral
$\int_{{\bf r}^{\prime}}^{\bf r}
 {\bf A}(\mbox{{\mathversion{bold}${\xi}$}})\cdot
d\mbox{{\mathversion{bold}${\xi}$}}$, we must evaluate
 $\int_{{\bf r}^{\prime}}^{\bf r}[
 {\bf A}(\mbox{{\mathversion{bold}${\xi}$}})-(1/2){\bf B}\times
(\mbox{{\mathversion{bold}${\xi}$}}-{{\bf r}}^{\;\prime})]\cdot
 d\mbox{{\mathversion{bold}${\xi}$}}$.
\item Since we solved the Heisenberg equations for the gauge invariant
operators ${\bf R}_\bot$ and
$\mbox{{\mathversion{bold}${\pi}$}}_\bot$, our final result is
written for a generic gauge. Note that the gauge-independent and
gauge-dependent parts of the propagator are clearly separated. The
gauge fixing corresponds to choose a particular expression for
 $\bf A(\mbox{{\mathversion{bold}${\xi}$}})$. Besides, from
 (\ref{PropagadorFinalBUniforme}) we imediately obtain
 the transformation law for the propagator under a gauge
 transformation ${\bf A} \rightarrow {\bf A} +
  \mbox{{\mathversion{bold}${\nabla}$}}\Lambda$, namely,
\begin{equation}\label{GaugeTransformation}
 K({\bf r},{{\bf r}}^{\;\prime};\tau)\longmapsto
\mbox{\large $ e^{\frac{ie}{\hbar}\,\Lambda({\bf r})}$}\, K({\bf
r},{{\bf r}}^{\;\prime};\tau)\, \mbox{\large $
e^{-\frac{ie}{\hbar}\,\Lambda({{\bf r}}^{\;\prime})}$}\; .\nonumber
\end{equation}
Although this transformation law was obtained in a particular case,
it can be shown that it is quite general.
\item It is interesting to show how the energy spectrum (Landau
levels), with the corresponding degeneracy per unit area, can be
extracted from propagator (\ref{PropagadorFinalBUniforme}). With
this purpose, we recall that the partition function can be obtained
from the Feynman propagator by taking $\tau=-i\hbar\beta$, with
$\beta=1/(K_BT)$, and taking the spatial trace,
$$
Z(\beta) = \int_{-\infty}^\infty dx_1\int_{-\infty}^\infty dx_2\;
K({\bf r},{\bf r};-i\hbar\beta)\; .
$$
Substituting (\ref{PropagadorFinalBUniforme}) into last expression,
we get
$$
Z(\beta) = \int_{-\infty}^\infty dx_1\int_{-\infty}^\infty dx_2\;
{m\omega\over 2\pi\hbar\,\mbox{senh}(\hbar\beta\omega)}\; ,
$$
where we used the fact that $\sin(-i\theta)=-i\,\sinh\,\theta$.
Observe that the above result is divergent, since the area of the
${\cal OX}_1{\cal X}_2$ plane is infinite. This is a consequence of
the fact that each Landau level is infinitely degenerated, though
the degeneracy per unit area is finite. In order to proceed, let us
assume an area as big as we want, but finite. Adopting this kind or
regularization, we write
\begin{eqnarray}
\int_{-L/2}^{L/2} \!\! dx_1\!\!\int_{-L/2}^{L/2}\!\! dx_2\;
\!\!\!&K&\!\!\! ({\bf r},{\bf r};-i\hbar\beta)\approx
{L^2\, m\omega\over 2\pi\hbar\,\mbox{senh}(\hbar\beta\omega)}\nonumber\\
\nonumber\\
&=& {L^2\,  eB\over 2\pi\hbar\left(\mbox{\large $
e^{\hbar\beta\omega}$}-
\mbox{\large $ e^{-\hbar\beta\omega}$}\right)}\nonumber\\
\nonumber\\
&=& {L^2\,  eB\over 2\pi\hbar} {\mbox{\large $ e^{-{1\over
2}\hbar\beta\omega_c}$}\over \left(1-\mbox{\large $
e^{-\hbar\beta\omega_c}$}\right)}
\nonumber\\
\nonumber\\
&=& \sum_{n=0}^\infty {L^2\,  eB\over 2\pi\hbar} \mbox{\large $
e^{-\beta(n+{1\over2})\hbar\omega_c}$}\; ,\nonumber
\end{eqnarray}
where we denoted by $\omega_c = eB/2m$ the ciclotron frequency.
Comparing this result with that of a partition function whose energy
level $E_n$ has degeneracy $g_n$, given by
$$
Z(\beta) = \sum_n g_n\; \mbox{\large $ e^{-\beta E_n}$}\; ,
$$
we imediately identify the so called Landau leves and the
corresponding degeneracy per unit area,
\begin{equation}\label{niveisdeLandau}
E_n = \left( n+{1\over 2}\right)\hbar\omega_c\;\; ;\;\;
 {g_n\over A} = {eB\over 2\pi\hbar}\;\;\;\; (n=0,1,...)\;
 .\nonumber
\end{equation}
\end{enumerate}

\subsection{Charged harmonic oscillator in a uniform magnetic field}
\label{choumf}

In this section we consider a particle with mass $m$ and charge $e$
in the presence of a constant and uniform magnetic field $\textbf{B}
= B{\bf e_3}$ and submitted to a 2-dimensional isotropic harmonic
oscillator potential in the ${\cal OX}_1{\cal X}_2$ plane, with
natural frequency $\omega_{0}$. Using the same notation as before,
we can write the hamiltonian of the system in the form
\begin{equation}
\mathcal{H} = \mathcal{H}_{\bot} + \frac{P_3^2}{2m}, \label{Eq48}
\end{equation}
where
\begin{equation}
\mathcal{H}_{\bot} = \frac{{\pi}_{1}^{2}+{\pi}_{2}^{2}}{2m} +
 \frac{1}{2}m\omega_0^2\left(X_1^2 + X_2^2\right). \label{Eq49}
\end{equation}

As before, the Feynman propagator for this problem takes the form
$K(\textbf{x},\textbf{x}^{\prime};\tau)=K_{\bot}
(\textbf{r},\textbf{r}^{\prime};\tau)K_{3}^{(0)}(x_3,x_3^{\prime};\tau)$,
with $K_{3}^{(0)}(x_3,x_3^{\prime};\tau)$ given by equation
(\ref{Eq14}). The propagator in the  ${\cal OX}_1{\cal X}_2$-plane
satisfies the differential equation
(\ref{EqDiferencialPropagador2D}) and  will be determined by the
same used in the previous example.

Using hamiltonian (\ref{Eq49}) and the usual commutation relations
 the Heisenberg equations  are given by
\begin{eqnarray}
\frac{d\textbf{R}(\tau)}{d\tau}&=&\frac{\mbox{\mathversion{bold}${\Pi}$}(\tau)}{m} \,,  \label{Eq51} \\
\frac{d\mbox{\mathversion{bold}${\Pi}$}(\tau)}{d\tau} &=& 2\omega
 \mathbb{C}  \mbox{\mathversion{bold}${\Pi}$}(\tau) - m
\omega_{0}^{2}\textbf{R}(\tau)\ ,  \label{Eq52}
\end{eqnarray}
where we have used the matrix notation introduced in (\ref{Eq20})
and (\ref{Eq21}). Equation (\ref{Eq51}) is the same as
(\ref{EquacaoR}), but equation (\ref{Eq52}) contains an extra term
when compared to (\ref{EquacaoPi}). In order to decouple equations
(\ref{Eq51}) and (\ref{Eq52}), we differentiate (\ref{Eq51}) with
respect to $\tau$ and then use (\ref{Eq52}). This procedure leads to
the following uncoupled equation
\begin{eqnarray}
\frac{d^{2}\textbf{R}(\tau)}{d\tau^{2}}
 &-&
 2\omega\mathbb{C}\frac{d\textbf{R}(\tau)}{d\tau}+\omega_{0}^{2}\textbf{R}(\tau)=0
\label{Eq53}
\end{eqnarray}
 After solving this equation, $\textbf{R}(\tau)$ and
$\mbox{\mathversion{bold}${\Pi}$}(\tau)$ are constrained to satisfy
equations (\ref{Eq51}) and (\ref{Eq52}), respectively. A
straightforward algebra yields the solution
\begin{eqnarray}
\textbf{R}(\tau)&=&\mathbb{M}^{-}\textbf{R}(0)+\mathbb{N}\mbox{\mathversion{bold}${\Pi}$}(0)
\label{Eq56} \\
\mbox{\mathversion{bold}${\Pi}$}(\tau)
 &=&
 \mathbb{M}^{+}\mbox{\mathversion{bold}${\Pi}$}(0)-m^{2}\omega_{0}^{2}\mathbb{N}\textbf{R}(0)\
, \label{Eq57}
\end{eqnarray}
where we defined the matrices
\begin{eqnarray}
\mathbb{N}&=&\frac{\sin{(\Omega\tau)}}{m\Omega}e^{\omega\tau\mathbb{C}} \label{Eq58}\\
\mathbb{M}^{\pm}&=&e^{\omega\tau\mathbb{C}}\Big[\cos{(\Omega\tau)}\uma\pm
\frac{\omega}{\Omega}\sin{(\Omega\tau)}\mathbb{C}\Big]\ ,
\label{Eq59}
\end{eqnarray}
and frequency $\Omega = \sqrt{\omega^{2}+\omega_{0}^{2}}$. Using
 (\ref{Eq56}) and (\ref{Eq57}), we write
$\mbox{\mathversion{bold}${\Pi}$}(0)$ and
$\mbox{\mathversion{bold}${\Pi}$}(\tau)$ in terms of
$\textbf{R}(\tau)$ and $\textbf{R}(0)$,
\begin{eqnarray}
\!\!\!\!\!\!\!\mbox{\mathversion{bold}${\Pi}$}(0)
 \!\!&=&\!\!
 \mathbb{N}^{-1}\textbf{R}(\tau)-\mathbb{N}^{-1}\mathbb{M}^{-}\textbf{R}(0)\,
 ,
\label{Pi(0)OH+B} \\
\!\!\!\!\!\!\!\mbox{\mathversion{bold}${\Pi}$}(\tau)
 \!\!\!&=&\!\!
 \mathbb{M}^{+}\mathbb{N}^{-1}\textbf{R}(\tau) \! -\!\!
\Big[ \mathbb{M}^{+}\mathbb{N}^{-1}\mathbb{M}^{-} \!\!\!+\!
m^{2}\omega_{0}^{2}\mathbb{N}\Big]\! \textbf{R}(0).
\label{Pi(tau)OH+B}
\end{eqnarray}

Now, we must order appropriately the hamiltonian operator
$\mathcal{H}_{\bot}=\mbox{\mathversion{bold}${\Pi}$}^{T}(0)
\mbox{\mathversion{bold}${\Pi}$}(0)/(2m)+m\omega_{0}^{2}\textbf{R}^{T}(0)\textbf{R}(0)/2$,
which, with the aid of equation (\ref{Pi(0)OH+B}), can be written as
\begin{eqnarray}
\mathcal{H}_{\bot}&=& \Big[\textbf{R}^{T}(\tau)(\mathbb{N}^{-1})^T -
\textbf{R}^{T}(0)(\mathbb{M}^{-})^T (\mathbb{N}^{-1})^T \Big] \cr
&&\times \Big[\mathbb{N}^{-1} \textbf{R}(\tau) -
\mathbb{N}^{-1}\mathbb{M}^{-}  \textbf{R}(0)\Big] +m\omega_0^2
\textbf{R}^{T}(0)\textbf{R}(0) \cr\cr
&=&\frac{m\Omega^{2}}{2\sin^{2}{(\Omega\tau)}}
\Big[\textbf{R}^{T}(\tau) - \textbf{R}^{T}(0)(\mathbb{M}^{-})^T
\Big] \cr &&\times \Big[\textbf{R}(\tau) - \mathbb{M}^{-}
\textbf{R}(0)\Big] +m\omega_0^2 \textbf{R}^{T}(0)\textbf{R}(0)
\cr\cr &=&\frac{m\Omega^{2}}{2\sin^{2}{(\Omega\tau)}}
\Big[\textbf{R}^{T}(\tau) \textbf{R}(\tau) - \textbf{R}^{T}(\tau)
\mathbb{M}^{-}\textbf{R}(0) \cr &&-
\textbf{R}^{T}(0)(\mathbb{M}^{-})^T \textbf{R}^{T}(\tau) +
\textbf{R}^{T}(0)(\mathbb{M}^{-})^T \mathbb{M}^{-} \textbf{R}(0)
\Big] \cr &&\;\;+m\omega_0^2 \textbf{R}^{T}(0)\textbf{R}(0) \cr\cr
&=&\frac{m\Omega^{2}}{2\sin^{2}{(\Omega\tau)}}\Big[\textbf{R}^{2}(\tau)
  -\textbf{R}^{T}(\tau)\mathbb{M}^{-}\textbf{R}(0)
\cr
&& \qquad\qquad -\textbf{R}^{T}(0)(\mathbb{M}^{-})^{T}\textbf{R}(\tau) +\textbf{R}^{2}(0)\Big]\ ,
\label{Eq63}
\end{eqnarray}
where superscript $T$ means transpose and we have used the
properties of the matrices $\mathbb{N}$ and $\mathbb{M}^{-}$ given
by (\ref{Eq58}) and (\ref{Eq59}). In order to get the right time
ordering, observe first that
$$
\textbf{R}^T(0)(\mathbb{M}^{-})^T \textbf{R}(\tau) =
\textbf{R}^T(\tau) \mathbb{M}^{-}\textbf{R}(0) + \Big[\left(
\mathbb{M}^{-} \textbf{R}(0)\right)_{i},\textbf{X}_{i}(\tau)\Big]
\,,
$$
where
$$
\Big[\left( \mathbb{M}^{-}
\textbf{R}(0)\right)_{i},\textbf{X}_{i}(\tau)\Big]
 = i\hbar \mbox{Tr}\Big[\mathbb{N}(\mathbb{M}^{-})^{T} \Big]
= \frac{i\hbar}{m\Omega} \sin{(2\Omega\tau)}\, .
$$
Using the last two equations into (\ref{Eq63}) we rewrite the
hamiltonian in the desired ordered form, namely,
\begin{eqnarray}\label{Hordenada1}
\mathcal{H}_{\bot}&=&\frac{m\Omega^{2}}{2\sin^{2}{(\Omega\tau)}}\Big[\textbf{R}^{2}(\tau)+\textbf{R}^{2}(0) - 2\textbf{R}^{T}(\tau)\mathbb{M}^{-}\textbf{R}(0)   \nonumber \\
&&  \qquad \qquad - \frac{i\hbar}{m\Omega} \sin{(2\Omega\tau)}
\Big]\ .
\end{eqnarray}
For future convenience, let us define
\begin{eqnarray}
U(\tau)&=&\cos{(\omega\tau)}\cos{(\Omega\tau)}
 + \frac{\omega}{\Omega}\sin{(\omega\tau)}\sin{(\Omega\tau)}\, , \label{Eq67} \\
V(\tau)&=&\sin{(\omega\tau)}\cos{(\Omega\tau)}
 - \frac{\omega}{\Omega}\cos{(\omega\tau)}\sin{(\Omega\tau)}
\label{Eq68}
\end{eqnarray}
and write matrix $\mathbb{M}^{-}$, defined in (\ref{Eq59}), in the
form
\begin{equation}
\mathbb{M}^{-}=U(\tau)\uma + V(\tau)\mathbb{C}. \label{Eq66}
\end{equation}
Substituting (\ref{Eq66}) in (\ref{Hordenada1}) we have
\begin{eqnarray}
\mathcal{H}_{\bot}&=&\frac{m\Omega^{2}}{2\sin^{2}{(\Omega\tau)}}\Big[\textbf{R}^{2}(\tau)+\textbf{R}^{2}(0)
 - 2U(\tau)\textbf{R}^{T}(\tau)\textbf{R}(0)   \nonumber \\
  &-&   2V(\tau)\textbf{R}^{T}(\tau)\mathbb{C}\textbf{R}(0)-\frac{i\hbar}{m\Omega}
  \sin{(2\Omega\tau)}\Big]\,.
\label{Eq69}
\end{eqnarray}
The next step is to compute the classical function
 $F({\bf r},{\bf r^{\prime}};\tau)$. Using the following identities
\begin{eqnarray}
\frac{\Omega U(\tau)}{\sin^{2}{(\Omega\tau)}}
 &=&
 - \frac{d}{d\tau} \Big[\frac{\cos{(\omega\tau)}} {\sin{(\Omega\tau)}}\Big]\, , \label{Eq70} \\
\frac{\Omega V(\tau)}{\sin^{2}{(\Omega\tau)}}
 &=&
 - \frac{d}{d\tau}\Big[\frac{\sin{(\omega\tau)}} {\sin{(\Omega\tau)}}\Big],
\label{Eq71}
\end{eqnarray}
into (\ref{Eq69}), we write $F(\textbf{r},\textbf{r}^{\prime};\tau)$
in the convenient form
\begin{eqnarray}
\!\!\!\! F(\textbf{r},\textbf{r}^{\prime};\tau)
 \!\!&=&\!\!
 \frac{m\Omega^{2}}{2}(\textbf{r}^{2}+\textbf{r}^{\prime^{2}})\mbox{csc}(\Omega\tau)^{2}
\!\!+ m\Omega\textbf{r}\cdot\textbf{r}^{\prime}\! \frac{d}{d\tau}
\! \Big[ \frac{\cos{(\omega\tau)}} {\sin{(\Omega\tau)}}\Big] \nonumber \\
&+& \; m\Omega\textbf{r}\cdot \mathbb{C}\textbf{r}^{\prime}
 \frac{d}{d\tau} \Big[ \frac{\sin{(\omega\tau)}}
 {\sin{(\Omega\tau)}}\Big]-i \hbar \Omega
 \frac{\cos{(\Omega\tau)}}{\sin{(\Omega\tau)}}\, .
\label{Eq72}
\end{eqnarray}
Inserting this result into the differential equation
$$
i\hbar\frac{\partial}{\partial\tau}\langle{\bf r},\tau
 \vert{\bf r}^{\prime},0\rangle =
 F(\textbf{r},\textbf{r}^{\prime};\tau)
\langle{\bf r},\tau
 \vert{\bf r}^{\prime},0\rangle\, ,
 $$
 and integrating in $\tau$, we obtain
\begin{eqnarray}
\langle\textbf{r},\tau \vert \textbf{r}^{\prime}\!\!\!\!\!
&,&\!\!\!\!\! 0\rangle = \frac{C(\textbf{r},\textbf{r}^{\prime})}
 {\sin{(\Omega\tau)}} \mbox{exp} \left\lbrace
  \frac{im\Omega}{2\hbar}\Big[ (\textbf{r}^{2}
  + \textbf{r}^{\prime^{2}})\cot{(\Omega\tau)}   \right.  \nonumber \\
  &-&
  \left. 2 \left( \textbf{r}\cdot\textbf{r}^{\prime}
  \frac{\cos{(\omega\tau)}}{\sin{(\Omega\tau)}}
  + \textbf{r}\cdot \mathbb{C}\textbf{r}^{\prime}
  \frac{\sin{(\omega\tau)}}{\sin{(\Omega\tau)}} \right)  \Big] \right\rbrace.
\label{ProtoPropagadorOH+B}
\end{eqnarray}
where $C({\bf r},{\bf r}^{\;\prime})$ is an arbitrary integration
constantto be determined by conditions (\ref{Eq30}), (\ref{Eq31})
and (\ref{Eq32}).  Using (\ref{Pi(tau)OH+B}) we can calculate the
l.h.s. of condition (\ref{Eq30}),
\begin{eqnarray}
\langle{\bf r}, \!\!\!\!&\tau&\!\!\!\! \vert
 \pi_j(\tau)\vert{\bf r}^{\prime},0\rangle
 =
 \frac{m\Omega}{\sin{(\Omega\tau)}}\Big\{ \cos{(\Omega\tau)}x_j-\cos{(\omega\tau)}x^{\prime}_j \nonumber \\
&+&
 \Big[\frac{\omega}{\Omega}\sin{(\Omega\tau)} x_k-\sin{(\omega\tau)}x^{\prime}_{k}\Big]
  \epsilon_{jk3} \Big\}\langle{\bf r},\tau\vert{\bf r}^{\prime},0\rangle, \label{Eq74}
\end{eqnarray}
and using (\ref{Pi(0)OH+B}) we get the l.h.s. of condition
(\ref{Eq31}),
\begin{eqnarray}
\langle{\bf r}, \!\!\!\!&\tau&\!\!\!\! \vert\pi_j(0)\vert{\bf
r}^{\prime},0\rangle = \frac{m\Omega}{\sin{(\Omega\tau)}}
 \Big\{\cos{(\omega\tau)}x_j-\cos{(\Omega\tau)}x^{\prime}_j \nonumber \\
&+&  \Big[\frac{\omega}{\Omega}\sin{(\Omega\tau)} x^{\prime}_k-\sin{(\omega\tau)}x_{k}\Big]\epsilon_{jk3} \Big\}
 \langle{\bf r},\tau\vert{\bf r}^{\prime},0\rangle.
\label{Eq75}
\end{eqnarray}
With the help of the simple identities
\begin{eqnarray}
\frac{\partial}{\partial x_j}(\textbf{r}^{2}+\textbf{r}^{\prime^{2}})=2x_j \hspace{0.198cm}
&;&
\hspace{0.198cm} \frac{\partial}{\partial x^{\prime}_j}(\textbf{r}^{2}
 + \textbf{r}^{\prime^{2}})=2x^{\prime}_j \nonumber \\
\frac{\partial}{\partial x_j}\textbf{r}\cdot\textbf{r}^{\prime}=x^{\prime}_j  \hspace{0.658cm}
&;&
 \hspace{0.658cm} \frac{\partial}{\partial x^{\prime}_j}\textbf{r}\cdot\textbf{r}^{\prime}= x_j \nonumber \\
 \frac{\partial}{\partial x_j}\textbf{r}\cdot
 \mathbb{C}\textbf{r}^{\prime}=\epsilon_{jk3}x^{\prime}_k  \hspace{0.2295cm}
&;&
\hspace{0.2295cm} \frac{\partial}{\partial x^{\prime}_j}\textbf{r}\cdot
 \mathbb{C}\textbf{r}^{\prime}= -\epsilon_{jk3}x_k. \nonumber
\end{eqnarray}
and also using equation (\ref{ProtoPropagadorOH+B}),  we are able to
compute the right hand sides of conditions (\ref{Eq30}) and
(\ref{Eq31}), which are given, respectively, by
\begin{eqnarray}
\Big\{
 \!\!\!\!&-&\!\!\!\!
 \frac{i\hbar}{C(\textbf{r},\textbf{r}^{\prime})}\frac{\partial
C(\textbf{r},\textbf{r}^{\prime})} {\partial x_j}
 + m\Omega\frac{\cos{(\Omega\tau)}}{\sin{(\Omega\tau)}}x_j
 -  m\Omega\frac{\cos{(\omega\tau)}}{\sin{(\Omega\tau)}}x^{\prime}_j
 \nonumber \\
 &-&
 m\Omega\frac{\sin{(\omega\tau)}}{\sin{(\Omega\tau)}}\epsilon_{jk3}x^{\prime}_k
 - e A_j(\textbf{r}) \Big\} \langle\textbf{r},\tau|\textbf{r}^{\prime},0\rangle
\label{Eq77}
\end{eqnarray}
and
\begin{eqnarray}
\Big\{
 \!\!\!\!\!&{\,}&\!\!\!\!\!
 \frac{i\hbar}{C(\textbf{r},\textbf{r}^{\prime})}
 \frac{\partial C(\textbf{r},\textbf{r}^{\prime})} {\partial x^{\prime}_j}
  m\Omega\frac{\cos{(\Omega\tau)}}{\sin{(\Omega\tau)}}x^{\prime}_j
 +  m\Omega\frac{\cos{(\omega\tau)}}{\sin{(\Omega\tau)}}x_j
  \nonumber\\
 &-&
 m\Omega\frac{\sin{(\omega\tau)}}{\sin{(\Omega\tau)}}\epsilon_{jk3}x_k
 -e A_j(\textbf{r}^{\prime})\Big\} \langle\textbf{r},\tau|\textbf{r}^{\prime},0\rangle.
\label{Eq78}
\end{eqnarray}
Equating (\ref{Eq74}) and (\ref{Eq77}), and also
 (\ref{Eq75}) and (\ref{Eq78})), we get the system of differential
 equations for $C(\textbf{r},\textbf{r}^{\prime})$
\begin{eqnarray}
i\hbar \frac{\partial C(\textbf{r},\textbf{r}^{\prime})}{\partial
 x_j}
 &+&
 e\Big[A_j(\textbf{r})+\frac{F_{jk}}{2}x_k \Big] C(\textbf{r},\textbf{r}^{\prime})=0\, , \label{Eq79} \\
i\hbar \frac{\partial C(\textbf{r},\textbf{r}^{\prime})}{\partial x^{\prime}_j}
 &-&
 e\Big[A_j(\textbf{r}^{\prime})+\frac{F_{jk}}{2}x^{\prime}_k \Big] C(\textbf{r},\textbf{r}^{\prime})=0.
\label{Eq80}
\end{eqnarray}
Proceeding as in the previous example, we first integrate
(\ref{Eq79}). With this goal, we multiply it by $dx_j$, sum in $j$
and integrate it to obtain
\begin{equation}
C({\bf r},{\bf r}^{\;\prime})= C({{\bf r}}^{\;\prime},{\bf
r}^{\;\prime})
 \exp\left\lbrace {ie\over \hbar}_{\;\;\Gamma}\!\!
\int_{{\bf r}^{\prime}}^{{\bf r}}
\Big[ A_j(\mbox{\mathversion{bold}${\xi}$})+{F_{jk}\over 2}\xi_k\Big] d\xi_j \right\rbrace\, , \\
\label{CrrLinha1}
\end{equation}
where the path of integration $\Gamma$ will be specified in a
moment. Inserting expression (\ref{CrrLinha1}) into the second
differential equation (\ref{Eq80}), we get
$$
\frac{\partial}{\partial x_j^{\prime}}
 C({{\bf r}}^{\;\prime},{{\bf r}}^{\;\prime}) = 0
 \;\;\Longrightarrow\;\;
 C({{\bf r}}^{\;\prime},{{\bf r}}^{\;\prime}) =C_0\, ,
$$
where $C_0$ is a constant independent of ${{\bf r}}^{\;\prime}$,
 so that equation (\ref{CrrLinha1}) can be cast, after some convenient
 rearrangements, into the form
\begin{equation}
C({\bf r},{\bf r}^{\prime})=C_{0}\exp{\left\lbrace{ie\over\hbar}
 _{\;\;\Gamma}\!\!
\int_{{\bf r}^{\prime}}^{{\bf r}}
\Big[\textbf{A}(\mbox{\mathversion{bold}${\xi}$})
 - \frac{1}{2}\textbf{B}\times\mbox{\mathversion{bold}${\xi}$}\Big]\cdot
d\mbox{\mathversion{bold}${\xi}$}\right\rbrace }.
 \label{CrrLinha2}
\end{equation}
Note that the integrand has a vanishing curl  so that we can choose
the path of integration $\Gamma$ at our will. Choosing, as before,
the straight line between ${\bf r}^{\prime}$ and ${\bf r}$, it can
be  shown that
\begin{equation}
\int_{\textbf{r}^{\prime}}^{\textbf{r}}
 \Big[ \textbf{A}(\mbox{\mathversion{bold}${\xi}$})-\frac{\textbf{B}}{2}
 \times\mbox{\mathversion{bold}${\xi}$}\Big]
 \cdot d\mbox{\mathversion{bold}${\xi}$}=\int_{\textbf{r}^{\prime}} ^{\textbf{r}}
 \textbf{A}(\mbox{\mathversion{bold}${\xi}$})\cdot
 d\mbox{\mathversion{bold}${\xi}$}
 + \frac{1}{2}B\textbf{r}\cdot\mathbb{C}\textbf{r}^{\prime}\, ,
 \label{IntegralLinha1}
\end{equation}
where, for simplicity of notation, we omitted the symbol
$\Gamma_{sl}$ indicating that the line integral must be done along a
straight line. From equations (\ref{CrrLinha1}), (\ref{CrrLinha2}) e
(\ref{IntegralLinha1}), we get
\begin{equation}
C({\bf r},{\bf r}^{\prime}) =
 C_{0}\exp{\left\lbrace{ie\over\hbar}\int_{\atop{{{\bf
r}}^{\prime}}}^{{\bf r}}
\textbf{A}(\mbox{\mathversion{bold}${\xi}$})\cdot
d\mbox{\mathversion{bold}${\xi}$}\right\rbrace }
\exp{\left\lbrace{im\omega\over\hbar}\textbf{r}
 \cdot\mathbb{C}\textbf{r}^{\prime}\right\rbrace} ,
 \label{Eq86}
\end{equation}
which substituted back into equation (\ref{ProtoPropagadorOH+B})
yields
\begin{eqnarray}
\langle\textbf{r},\tau|\textbf{r}^{\prime},0\rangle
 = \frac{C_{0}}{\sin{(\Omega\tau)}}\exp{\left\lbrace{ie\over\hbar}
 \int_{\atop{{{\bf r}}^{\prime}}}^{{\bf r}} \textbf{A}(\mbox{\mathversion{bold}${\xi}$})
 \cdot d\mbox{\mathversion{bold}${\xi}$}\right\rbrace } \nonumber \\
\mbox{exp} \Big\{  \frac{im\Omega}{2\hbar\sin{(\Omega\tau)}}
 \Big\{(\textbf{r}^{2} +\textbf{r} ^{\prime^{2}}) \cos{(\Omega\tau)}
 \nonumber \\
- 2\textbf{r} \cdot\textbf{r}^{\prime}\cos{(\omega\tau)}
-  2\Big[\sin{(\omega\tau)}-\frac{\omega}{\Omega}\sin{\Omega\tau} \Big]
 \textbf{r}\mathbb{C}\textbf{r}^{\prime}  \Big\} \Big\}
\label{Eq87}
\end{eqnarray}
The initial condition implies $C_0=m\Omega/(2\pi i\hbar)$. Hence,
the desired Feynman propagator is finally given by
\begin{widetext}
\begin{eqnarray}
K({\bf x},{{\bf x}}^{\prime};\tau)
 &=&
K_{\bot}
(\textbf{r},\textbf{r}^{\prime};\tau)K_{3}^{(0)}(x_3,x_3^{\prime};\tau)\nonumber\\
&=&
 \frac{m\Omega}{2\,\pi\, i\, \hbar\,\sin{(\Omega\tau)}}
 \sqrt{{m\over 2\pi i\hbar\tau}} \exp{\left\lbrace{ie\over\hbar}
 \int_{\atop{{{\bf r}}^{\prime}}}^{{\bf r}} \textbf{A}(\mbox{\mathversion{bold}${\xi}$})
 \cdot d\mbox{\mathversion{bold}${\xi}$}\right\rbrace } \mbox{exp} \left\lbrace
 \frac{im\Omega}{2\hbar\sin{(\Omega\tau)}} \left\lbrace \cos{(\Omega\tau)}
 (\textbf{r}^{2} +\textbf{r} ^{\prime^{2}})  \right. \right. \nonumber \\
 &{\;}& \, -\;\;
  \left. \left. 2 \cos{(\omega\tau)} \textbf{r} \cdot\textbf{r}^{\prime} -
  2 \Big[\sin{(\omega\tau)}-\frac{\omega}{\Omega}\sin{(\Omega\tau)} \Big]
  \textbf{r}\cdot\mathbb{C}\textbf{r}^{\prime}  \right\rbrace \right\rbrace
  \exp\biggl\{{im\over 2\hbar}{\left(x_3 - x_3^\prime\right)^2\over \tau}\biggr\}\, ,
\label{Eq90}
\end{eqnarray}
\end{widetext}
where we brought back the free part of the propagator corresponding
to the movement along the ${\cal OX}_3$ direction. Of course, for
$\omega_0=0$ we reobtain the propagator found in our first example
and for ${\bf B}={\bf 0}$ we reobtain the propagator for a
bidimensional oscillator in the ${\cal OX}_1{\cal X}_2$ plane
multiplied by a free propagator in the ${\cal OX}_3$ direction, as
can be easily checked.

Regarding the gauge dependence of the propagator, the same comments
done before are still valid here, namely, the above expression is
written for a generic gauge, the transformation law for the
propagator under a gauge transformation is the same as before, etc.
We finish this section, extracting from the previous propagator, the
corresponding energy spectrum. With this purpose, we first compute
the trace of the propagator,
\begin{widetext}
\begin{eqnarray}
\int_{-\infty}^\infty\!\!\! dx_1\!\!
 \int_{-\infty}^\infty\!\!\! dx_2 \,
 K_\perp^{\,\prime}(x_1,x_1,x_2,x_2;\tau) &=& {m\Omega\over 2\pi
i\hbar\,\sin(\Omega\tau) }
 \int_{-\infty}^\infty\!\!\! dx_1
 \!\!\int_{-\infty}^\infty\!\!\! dx_2
 \exp\biggl\{ {im\Omega\over
2\hbar\,\sin(\Omega\tau)}\left[
2\Bigl(\mbox{cos}(\Omega\tau)-\mbox{cos}(\omega\tau)\Bigr)(x_1^2+x_2^2)\right]
\biggr\} \nonumber\\
 &=& {1\over 2[\mbox{cos}(\Omega\tau)-\mbox{cos}(\omega\tau)]}\; ,
\end{eqnarray}
\end{widetext}
where we used the well known result for the Fresnel integral. Using
now the identity
$$
\cos(\Omega\tau)-\cos(\omega\tau) = -2\,\sin[(\Omega+\omega)\tau/2]
\,\sin[(\Omega-\omega)\tau/2)]\, ,
$$
we get for the corresponding energy Green function
\begin{eqnarray}\label{FGreenG(E)}
\!\!&{\cal G}&\!\!\!(E) =\! -i\!\!\int_0^\infty\!\!\! d\tau\,
e^{{i\over\hbar}E\tau} \!\!\int_{-\infty}^\infty\, \!\!\!\!
dx_1\!\!\int_{-\infty}^\infty\!\!\!\! dx_2\,
K_\perp^{\,\prime}(x_1,x_1,x_2,x_2;\tau)\cr\cr
&=&{i\over4}\int_0^\infty d\tau {e^{{i\over\hbar}E\tau}\over
\,\mbox{sen}({\Omega\tau\over2}\tau)\,\mbox{sen}({\Omega-\omega\over2}\tau)}\cr\cr
&=&-i\!\!\int_0^\infty\!\!\!\! d\tau\,
e^{{i\over\hbar}E\tau}\left(\sum_{l=0}^\infty
e^{-(l+{1\over2})(\Omega+\omega)\tau}\!\!\right)\!\!
\left(\sum_{n=0}^\infty
e^{-i(n+{1\over2}(\Omega-\omega)\tau}\!\!\right) ,\nonumber
\end{eqnarray}
where is tacitly assumed that $E\rightarrow E-i\varepsilon$ and we
also used that (with the assumption $\nu\rightarrow \nu-i\epsilon$)
$$
{1\over\,\mbox{sen}({\nu\over2}\tau)}=2i\sum_{n-0}^\infty
e^{-i(n+{1\over2}) \nu\tau}\; .
$$
Changing the order of integration and summations, and integrating in
$\tau$, we finally obtain
\begin{equation}
{\cal G}(E)=\sum_{l,n=0}^\infty{1\over E-E_{nl}}\; ,
\end{equation}
where the poles of ${\cal G}(E)$, which give the desired energy
levels, are identified as
 \begin{equation}
 E_{nl}=(l+n+1)\hbar\Omega+(l-n)\hbar\omega\, ,\;\;\; (l,n = 0,1,...)\; .
 \end{equation}
The Landau levels can be reobtained from the previous result by
simply taking the limit $\omega_0\rightarrow 0$:
\begin{equation}
E_{nl}\longrightarrow(2l+1)\hbar\omega
 = (l+{1\over2})\hbar\omega_c\; ,
\end{equation}
with $l=0,1,...$ and $\omega_c = eB/m$, in agreement to the result
we had already obtained before.


${\;}$
\section{Final Remarks}\label{SectionFinalRemarks}

In this paper we reconsidered, in the context of Schwinger's method,
the Feynman propagators of two well known problems, namely, a
charged particle under the influence of a constant and uniform
magnetic field (Landau problem) and the same problem in which we
added a bidimensional harmonic oscillator potential. Although these
problems had already been treated from the point of view of
Schwinger's action principle, the novelty of our work relies on the
fact that we solved the Heisenberg equations for gauge invariant
operators. This procedure has some nice properties, as for instance:
\linebreak
 {\it (i)} the Feynman propagator is obtained in a generic
gauge; {\it (ii)} the gauge-dependent and gauge-independent parts of
the propagator appear clearly separated and {\it (iii)} the
transformation law for the propagator under gauge transformation can
be readly obtained. Besides, we adopted a matrix notation which can
be straightforwardly generalized to cases of relativistic charged
particles in the presence of constant electromagnetic fields and a
plane wave electromagnetic field, treated by Schwinger
\cite{Schwinger1951}. For completeness, we showed explicitly how one
can obtain the energy spectrum directly from que Feynman propagator.
In the Landau problem, we obtained the (infinitely degenerated)
Landau levels with the corresponding degeneracy per unit area. For
the case where we included the bidimensional harmonic potential,
 we obtained the energy spectrum after identifying the poles of the
 corresponding energy Green function. We hope that this pedagogical
 paper may be useful for undergraduate as well as graduate students and that these two simple
 examples may enlarge the (up to now) small list of non-relativistic problems
 that have been treated by such a powerful and elegant method.

\section*{Acknowledgments}

F.A. Barone, H. Boschi-Filho and C. Farina would like to thank
Professor Marvin Goldberger for a private communication and for
kindly sending his lecture notes on quantum mechanics where this
method was explicitly used. We would like to thank CNPq and Fapesp
(brazilian agencies) for partial financial support.

\end{document}